\input{aipcheck}
\documentclass[
final            
  ] {aipproc}
\usepackage{amsbsy}
\layoutstyle{6x9}

\def\nnb{{\bar{\mathrm{n}}\mathrm{n}}}
\def\LLb{{\overline{\Lambda}\Lambda}}
\def\LSb{{\overline{\Lambda}\Sigma}}
\def\SSb{{\overline{\Sigma}\Sigma}}
\def\Y{{\mathrm{Y}}}
\def\YYb{{\overline{\Y}\Y}}
\def\XXb{{\overline{\Xi}\Xi}}
\def\YYpb{{\overline{\Y}\Y'}}
\def\ppLL{{\bar{\mathrm{p}}\mathrm{p}\to\overline{\Lambda}\Lambda}}
\def\ppLS{{\bar{\mathrm{p}}\mathrm{p}\to\overline{\Lambda}\Sigma}}

\def\ppXX{{\bar{\mathrm{p}}\mathrm{p}\to\overline{\Xi}\Xi}}
\def\ppYY{{\bar{\mathrm{p}}\mathrm{p}\to\overline{\Y}\Y'}}
\def\ppnn{{\bar{\mathrm{p}}\mathrm{p}\to\bar{\mathrm{n}}\mathrm{n}}}
\def\K{{\mathrm{K}}}
\def\ssb{{\bar{s}s}}
\def\qqb{{\bar{q}q}}
\def\uub{{\bar{u}u}}
\def\ddb{{\bar{d}d}}

\begin{document} 
 \title{Dynamics of hyperon--antihyperon production
 \footnote{Invited talk at  LEAP 05, May 16-22, 2005, to appear in the  Proceedings}%
$^{\; ,}$%
\footnote{Preprint \# LPSC 05-71,\qquad ArXiv:nucl-th/0508061}} 
 \author{Jean-Marc
Richard}{address={%
Universit\'e Joseph Fourier-CNRS-IN2P3, 
53, av. des Martyrs, 38026 Grenoble cedex, France}} 
 \begin{abstract} 
 The dynamics of
the strangeness-exchange reactions $\ppLL$, $\LSb$ or $\SSb$ is discussed,
either at the hadron level, in terms of strange-meson exchange, or at the quark
level, in terms of quark--antiquark creation or annihilation.
 \end{abstract} 

\maketitle
 \section{Introduction}
The strangeness-exchange reactions $\ppYY$,
where $\Y$ is an hyperon $\Lambda$ or $\Sigma$,  have been studied at LEAR by
the PS185 collaboration, at several values of the beam momentum. The last
runs benefitted from a transversely-polarized target, giving access to a variety
of spin observables.

The results of this experiment had a large impact on our community, as
seen from the impressive number of articles devoted to these reactions in
recent years. It is unfortunately impossible to account here for all
contributions. We refer to a recent review article \cite{Klempt:2002ap} for a
more comprehensive bibliography.

Schematically, there are two kinds of approaches. The first one describes
$\ppYY$ by suitable exchanges of mesons, generalizing what is done rather
successfully for the long- and medium-range part of the nucleon--nucleon and
antinucleon--nucleon interaction.

The second approach is guided by the phenomenology developed to describe the
hadron spectrum, the strong decays of hadron resonances, and the
Okubo--Zweig--Iizuka (OZI) rule in terms of constituent quarks.

Both pictures describe the main features of the  strangeness-exchange
reactions, provided they are supplemented by a realistic treatment of the
initial- and final-state interaction. However, they differ somehow on their
predictions for the spin observables.  
\section{Hadronic picture} 
%
%
In the conventional theory of nuclear forces,  the nucleon--nucleon
interaction is mediated by the exchange of mesons. Each exchange has a rather
clear signature, linked to the quantum numbers of the meson. For instance, a
pseudoscalar exchange gives a spin-spin and a tensor components, while the
exchange of vector mesons is the main source of spin-orbit forces. Isovector
exchanges give contributions of different sign and strength to the isospin
$I=1$ and  $I=0$ channels. This means that a detailed analysis of spin
observables and a comparison of proton--proton and neutron--proton data can
probe the meson-exchange models. Years of measurements with polarized nucleon
beams and targets have led to a good understanding of the long- and
medium-range parts of the nucleon--nucleon interaction.

 It is thus natural to extend this picture to other processes. Fermi and Yang
\cite{Fermi:1959sa}, in particular, pointed out that the meson-exchange
potential, in a given isospin channel, remains the same when going from the
nucleon--nucleon to the antinucleon--nucleon case if the exchanged meson has
$G$-parity $G=+1$, while it flips sign if $G=-1$. This $G$-parity
rule  transforms any theoretical model of nuclear forces into predictions for the long- and intermediate-range parts of the antinucleon--nucleon interaction.

Building an antinucleon--nucleon potential starting from meson exchanges was attempted by several authors, such as Bryan and Philips, Dover and
Richard, Kohno and Weise, etc. For references, see, e.g., \cite{Klempt:2002ap}.
This might have been sometimes considered with skepticism, as the
cross-sections are dominated by a strong annihilation acting up to about $1\;$fm. However, the analyzing power and the energy-shifts of the P-state levels of protonium are more sensitive to the
external part of the interaction, and are well reproduced by the meson-exchange
models.

Of particular interest is the charge-exchange reaction $\ppnn$. The isospin
 content of its amplitude, which reads 
  \begin{equation}
 \mathcal{M}(\ppnn)\propto  \mathcal{M}(I=1)- \mathcal{M}(I=0)~, \end{equation}
 indicates a likely cancellation of most annihilation components, thus
 enhancing the role of the exchange of isovector mesons such as $\pi$ and
 $\rho$. This was discussed, e.g., at the LEAR Workshop held at Erice in 1982
 \cite{LEAR82}.
 The measurements performed by the PS199 collaboration and its sequel
 confirmed the role of meson-exchange in $\ppnn$ at low
 energy  \cite{Klempt:2002ap}.

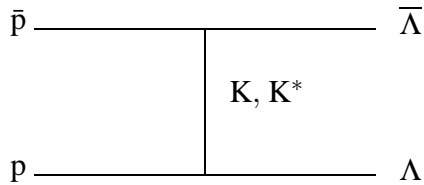
\begin{figure}[h] \label{fig:k-exc}
        {\tt    \setlength{\unitlength}{0.92pt}
\begin{picture}(240,85)(0,20)
\thinlines    \put(120,50){${\rm K},\,{\rm K}^*$}
              \put(190,18){$\Lambda$}
              \put(190,78){$\overline{\Lambda}$}
              \put(30,20){${\rm p}$}
              \put(30,80){$\bar{\rm p}$}
              \put(110,80){\line(0,-1){60}}
              \put(110,80){\line(1,0){70}}
              \put(110,20){\line(1,0){70}}
              \put(40,20){\line(1,0){70}}
              \put(40,80){\line(1,0){70}}
\end{picture}
}
\caption{Meson-exchange dynamics for $\ppLL$.}
\end{figure}

 Is is thus tempting to generalize this approach to include
strangeness-exchange, by replacing  $\nnb$ by $\LLb$, or more generally $\YYb$
or $\YYpb$, if $\Y$ denotes a strange baryon. Hence $\pi$ exchange becomes $\K$
exchange, $\rho$ is replaced by $\K^*$, etc. 
%
%
The dynamics is pictured by the diagram of Fig.~\ref{fig:k-exc}.
A strong tensor force is predicted in this approach, as for $\ppnn$. The latter
was unfortunately never probed in detail. Here, thanks to the information
provided by the weak decay of $\overline{\Lambda}$ and $\Lambda$ on the
final-state polarization, more observables are available.

Meson-exchange models of $\ppYY$ have been developed by the Bonn--J\"ulich
group, Furui and F\"assler, Tabakin et al., La France et al., among others
\cite{Klempt:2002ap}.
\section{Quark dynamics}
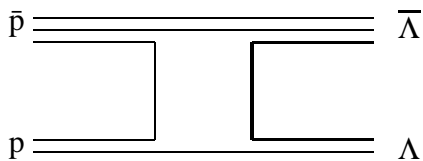
\begin{figure}[!hbtc]
        {\tt    \setlength{\unitlength}{0.92pt}
\begin{picture}(240,70)(0,10)
\linethickness{.9pt}   
              \put(140,60){\line(0,-1){40}}
              \put(140,20){\line(1,0){50}}
              \put(140,60){\line(1,0){50}}
\linethickness{.4pt}     
              \put(100,60){\line(0,-1){40}}
              \put(50,60){\line(1,0){50}}
              \put(50,20){\line(1,0){50}}
              \put(50,15){\line(1,0){140}}
              \put(50,65){\line(1,0){140}}
              \put(50,10){\line(1,0){140}}
              \put(50,70){\line(1,0){140}}
              \put(200,12){$\Lambda$}
              \put(200,62){$\overline{\Lambda}$}
              \put(40,15){${\rm p}$}
              \put(40,65){$\bar{\rm p}$}
              \end{picture}}
\caption{Simplest quark diagram  for $\ppLL$. The ordinary quarks or 
antiquarks are 
shown with thin lines, the strange ones by thick lines.\label{fig:quark1}}
\end{figure}

Since ${\mathrm K}$ or ${\mathrm K}^*$ mesons are relatively heavy,
the production of hyperons is a rather short-range process and,
instead of summing over all possible kaon excitations in the
$t$-channel, one might think of a simple quark process, as pictured in
Fig.~\ref{fig:quark1}: a pair of ordinary quarks annihilate and a pair 
of strange quarks is created.  Gluons are not shown, but are crucial 
to actually generate the process.
    
One should stress once more that these pictures are not Feynman diagrams of a
well-defined field theory. They simply describe the flow of flavor from initial
to final states. They have to be supplemented by models providing the wave
functions to be folded, and operators for quark--antiquark creation or
annihilation.  There are interesting differences her from an author to another.
See, e.g., the review by Alberg \cite{Alberg:2001fv} for a discussion and references.

In the phenomenology of hadron-resonance decay, it is often assumed that the
quark--antiquark pair is created with the quantum numbers of a vacuum, this is
the so-called ${}^3\mathrm{P}_0$ model \cite{LeYaouanc:1988fx}. Other choices
are, however, possible, such as ${}^3\mathrm{S}_1$, which would correspond to
the quantum numbers of a single gluon, or a superposition of several partial
waves.

\begin{figure}[h]\label{fig:quark2}
{\tt    \setlength{\unitlength}{0.92pt}
\begin{picture}(240,75)(0,5)
\linethickness{.9pt} 
              \put(150,60){\line(0,-1){40}}
              \put(50,20){\line(1,0){100}}
              \put(50,15){\line(1,0){140}}
              \put(150,60){\line(1,0){40}}
\linethickness{.4pt}   
              \put(50,65){\line(1,0){140}}
              \put(50,70){\line(1,0){140}}
              \put(50,10){\line(1,0){140}}
              \put(50,5){\line(1,0){140}}
              \put(50,25){\line(1,0){50}}
              \put(100,25){\line(0,1){35}}
              \put(50,60){\line(1,0){50}}
              \put(200,7){$\Lambda$}
              \put(200,62){$\overline{\Lambda}$}
              \put(40,15){${\rm p}$}
              \put(40,65){$\bar{\rm p}$}
            \end{picture}}
\caption{Possible sea-quark contribution to $\ppLL$.
The strange quarks or antiquarks are shown with thick lines.
}
\end{figure}
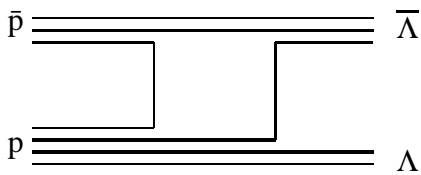

While PS185 was taking data and seeking further spin observables, new ideas
were developed, motivated by experiments on the structure functions of the
nucleon.  A $\ssb$ pair might be extracted from the nucleon or antinucleon sea
instead of being created during the reaction, as schematically pictured in
Fig.\ref{fig:quark2} and proposed  by Alberg
et al.~\cite{Alberg:1995zp}, who stressed that a similar mechanism could
produce an abundant violation of the OZI rule in annihilation.

\section{Models facing data}
Though rather different, the meson-exchange and the quark-inspired models gave
similar good account of the first PS185 data. In particular, the dominance of
the spin-triplet states over the spin-singlet state was well reproduced. In the
nuclear physics approach, this is due to the strong tensor force acting in
triplet states and vanishing in singlet.  In the quark model of Fig.~\ref{fig:quark1}, the light $qq$ and
$\bar{q}\bar{q}$ diquarks having spin 0, and the annihilating $\qqb$ and the
created $\ssb$ pairs spin 1, an overall spin triplet is required.

Then it was hoped that more refined spin observables would distinguish among the
models. Alberg et al.\ \cite{Alberg:1995zp} pointed out that the model of Fig.~\ref{fig:quark2}
implies large values for some spin observables, to the extent that the $\ssb$
pair in the initial proton or antiproton is polarized.
Holinde et al.\ \cite{Haidenbauer:1992gv} made predictions for the spin observables allowed by a
polarized target and reached the conclusion that the mechanisms of
Fig.~\ref{fig:k-exc} and Fig.~\ref{fig:quark1} give significantly different values for
some observables.

As shown at this conference, the data came eventually in between the predictions
of the meson-exchange and the quark-pair creation models. This means that other
effects have to be accounted for. In particular, our knowledge of the initial
state is limited, due to the lack of data on spin correlations in elastic
$\bar{\mathrm{p}}\mathrm{p}\to\bar{\mathrm{p}}\mathrm{p}$.

As for the final state, we have basically no information. This is, however, a
crucial ingredient for $\ppLL$. La France et al., for instance, had to enforce a
very strong annihilation in the $\LLb$ channel \cite{LaFrance:1988if}.
\boldmath\section{Production of $\Sigma$}\unboldmath
Unlike $\LLb$ which filters isospin $I=0$, the reaction $\ppLS$ (the
charge conjugate $\overline{\Lambda}\Sigma$ is implicitely implied) forces an
isospin $I=1$. This means, very likely, differences in the spin-dependence of the
initial-state interaction, with less tensor forces, and more spin-orbit. 
The final state $\LSb$ might also differ from $\LLb$, with perhaps a slightly weaker annihilation.
In a naive quark model, the
spin of $\Lambda$ is carried by the $s$ quark, while the spin of a $\Sigma$ is 
opposite to that of its constituent $s$ quark. Hence if the spin correlation in
the final state is attributed to a specific state of $\ssb$, it is translated in
different effects for $\LLb$ and $\LSb$.

There are essential differences for $\overline{\Sigma}{}^+\Sigma^-$ and 
$\overline{\Sigma}{}^-\Sigma^+$. In a Yukawa-type of model, the production of
the latter can be explained by the exchange of a neutral kaon, and thus the
cross-section should be of the same order of magnitude as for $\LLb$ and $\LSb$.
The production of $\overline{\Sigma}{}^+\Sigma^-$ requires an exotic exchange,
namely a mesonic system with one unit of strangeness and two units of charge. It
is thus expected to be suppressed. The same difference can be seen in the
quark-diagram approach: $\LLb$, $\LSb$ and $\overline{\Sigma}{}^-\Sigma^+$ can
be reached by annihilating a single $\uub$ or $\ddb$ pair replaced by an $\ssb$
one. For the final state $\overline{\Sigma}{}^+\Sigma^-$, the simple quark
diagram of Fig.~\ref{fig:quark1} does not operate: one needs more pairs created
or annihilated. Holinde et al.\ \cite{Haidenbauer:1992hv} found, however, that once all ingredients of their calculation are included, the $\overline{\Sigma}{}^+\Sigma^-$ production is not too much suppressed.
\boldmath\section{Production of $\Xi$}\unboldmath
An exotic  exchange is also required to produce a $\XXb$ pair. Hence, an anomalous production of $\XXb$ would indicate the possibility of mesonic resonances with strangeness $S=2$. 
There is a long tradition, indeed, coming from duality, to complement the direct investigation of exotic states by an indirect study, which consists of  looking at reactions whose tentative mechanism is the exchange of these states. See, e.g., Refs. \cite{Rosner:1968si,Nicolescu:1978ac}.

If the dynamics of $\ppXX$ is seen from the $s$-channel point of view, there are contributions of $\ppLL$ followed by $\LLb\to\XXb$. This led Haidenbauer et al.\ \cite{Haidenbauer:1992wi} to estimate the cross-section for $\overline{\Xi}{}^0\Xi^0$ to be significantly smaller than for $\overline{\Xi}{}^+\Xi^-$. The same hierarchy was found by Kroll et al.\  \cite{Kroll:1988cd} in their diquark model, opposite to the earlier estimate by Genz et al.\  \cite{Genz:1991jm}.
\section{Outlook}
The study of the production of hyperon--antihyperon pairs will hopefully be resumed at the future hadron
facilities which have been presented at this conference. The underlying physics is, indeed, related to several important questions of strong-interaction  physics, such as the decay of resonances, the violation of flavor SU(3) symmetry, or the validity of  the OZI rule.

The charm threshold can be envisaged. Here the contrast should be even more pronounced, between the highly non-perturbative background, and the hard process responsible for $\bar{c}c$ production.

The PS185 experiment clearly demonstrated how useful are spin observables to extract information about the underlying mechanisms. We are now convinced, that if new experiments are once designed to improve our knowledge of antiproton-induced reactions, polarization should be envisaged from the very beginning. If the striking spin effects seen by PS185 have something to do with $\ssb$ creation, this should also show up in other final states with hidden strangeness. Already, important values have been measured for the spin observables of $\bar{\mathrm p}\mathrm{p}\to \overline{\K}{\K}$ \cite{Hasan:1992sm}.

The search for baryonium states was one of the main motivation for building the LEAR facility at CERN and in the design of most LEAR experiments. PS185 was no exception. In the first data, something like a peak was seen in the cross-section of $\ppLL$ as a function of the incoming energy, but the peak was not confirmed in a more detailed scan of the energy-range just above threshold.

The question of baryonium is perhaps not definitely settled. It remains that a very strong attraction is observed between most baryon--antibaryon pairs. Recently, peaks have been seen in the spectrum  of baryon--antibaryon states produced in $\mathrm{J}/\psi$ decay or $\mathrm{B}$-meson decay. This is the subject of contributions to this conference, see, e.g. ,\cite{Wycech:2005wg}.

The availability of several spin observables for the reaction $\ppLL$ at the same energy stimulated a renewed interest on the methodology required to handle these data. Elchikh et al.\ \cite{Elchikh:2005}
addressed the question of testing the compatibility between various observables. Each of them is typically normalized to belong to the interval $[-1,+1]$, but two of them are often restricted to a sub-domain of the square $[-1,+1]^2$, three of them to a sub-volume of the cube $[-1,+1]^3$, etc. This is expressed by inequalities which have to be fulfilled by the experimental values extracted from independent measurements.

More ambitious is the attempt to perform an amplitude analysis of the data. This is implicit in the PS185 analysis \cite{Bassalleck:2002sd} and explicit in a recent  article by Bugg \cite{Bugg:2004rj}.
The possibility of extracting unambiguously the amplitudes of a $1/2+1/2\to 1/2+1/2$ reaction has been often debated. A recent contribution is \cite{Paschke:2000mx}.
In the case of the elastic antiproton--proton scattering, the debate remains open on whether or not the available data are sufficient to determine the five complex amplitudes or the spin-dependent components of the potential. In the case of $\ppLL$, the situation is clearly better since more observables have been measured, and some of these observables are found to be large (in absolute value), thus restricting the range allowed for the yet unknown observables. An amplitude analysis has, indeed, been proposed \cite{Bugg:2004rj}. Several resonances or resonance-like structures have been identified,  which confirm the richness of the $\ppLL$ dynamics, and call for further investigations.

\subsection*{Acknowledgments}
I would like to thank the organizers of LEAP05 for the stimulating atmosphere of the conference.

\end{document}